\documentclass[journal,twoside,web]{ieeecolor}
\usepackage{tuffc}
\usepackage{cite}
\usepackage{amsmath,amssymb,amsfonts}
\usepackage{algorithmic}
\usepackage{graphicx}
\usepackage{textcomp}
\usepackage{wrapfig,colortbl}
\usepackage{array}
\usepackage{multirow}

\definecolor{abstractbg}{rgb}{1,0.969,0.914}
\def\BibTeX{{\rm B\kern-.05em{\sc i\kern-.025em b}\kern-.08em
    T\kern-.1667em\lower.7ex\hbox{E}\kern-.125emX}}
\begin{document}
\title{
Volumetric Ultrasound via 3D Null Subtraction Imaging with Circular and Spiral Apertures
}
\author{Bingze Dai, \IEEEmembership{Student Member, IEEE}, Xi Zhang, \IEEEmembership{Student Member, IEEE}, and Wei-Ning Lee, \IEEEmembership{Senior Member, IEEE}
\thanks{This work was supported by the Hong Kong Research Grants Council General Research Fund (grant numbers: 17205022). (Corresponding author: Wei-Ning Lee.)}
\thanks{Bingze Dai, Xi Zhang, and Wei-Ning Lee are with Department of Electrical and Electronic Engineering, The University of Hong Kong, Hong Kong. (e-mail: bingzed, u3008996@connect.hku.hk, wnlee@hku.hk). Wei-Ning Lee is also with the School of Biomedical Engineering, The University of Hong Kong, Hong Kong}
}

\IEEEtitleabstractindextext{%
\fcolorbox{abstractbg}{abstractbg}{%
\begin{minipage}{\textwidth}\rightskip2em\leftskip\rightskip\bigskip
\begin{wrapfigure}{r}{0.4\textwidth}
    \vspace{-1em} 
    \hspace{-5em}
    \centering
    \includegraphics[width=2.9in]{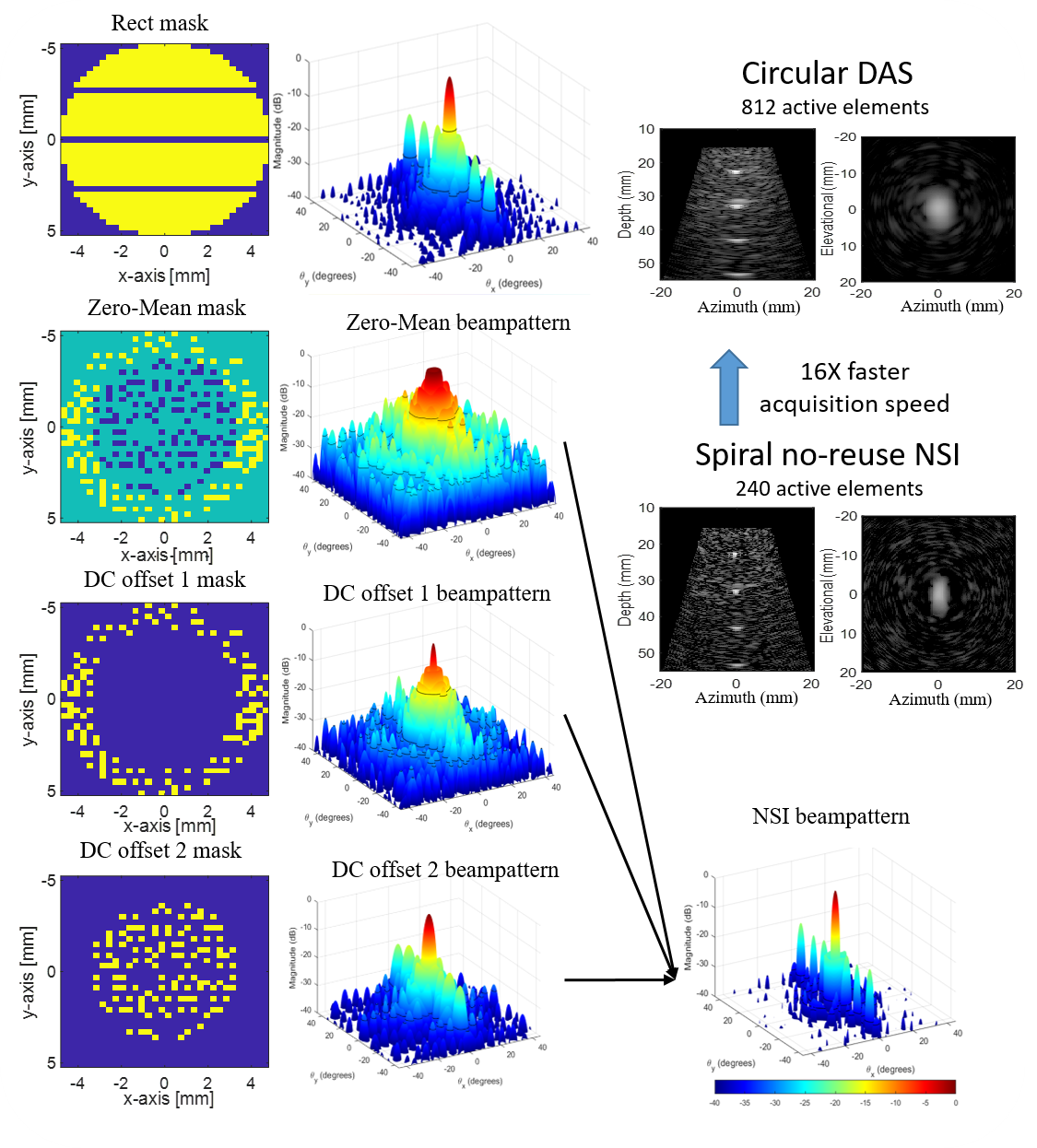} 
\end{wrapfigure}
\begin{abstract}

Volumetric ultrasound imaging faces a fundamental trade-off among image quality, frame rate, and hardware complexity. This study introduces three-dimensional Null Subtraction Imaging (3D NSI), a nonlinear beamforming framework that addresses this trade-off by combining computationally efficient null-subtraction process with multiplexing-aware sparse aperture designs on matrix arrays. We evaluate three apodization configurations: a fully addressed circular aperture and two Fermat's spiral sparse apertures. To overcome channel-sharing constraints common in matrix arrays multiplexed with low-channel-count ultrasound systems, we propose a spiral “no-reuse” apodization that enforces non-overlapping element sets across transmit–receive events. This design resolves multiplexing conflicts and enables up to a 16-fold increase in acquisition volume rate using only 240 active elements on a 1024-element probe. In computer simulations and tissue-mimicking phantom experiments, 3D NSI achieved an average improvement of 36\% in azimuthal and elevational resolutions, along with an approximately 20\% higher contrast ratio, compared to the conventional Delay-and-Sum (DAS) beamformer under matched transmit/receive configurations. 
When implemented with the spiral no-reuse aperture, the 3D NSI framework achieved over 1000 volumes per second with a computational load less than three times that of DAS, making it a practical solution for real-time 4D imaging.

\end{abstract}

\bigskip
\end{minipage}}}

\maketitle

\begin{table*}[!t]
\arrayrulecolor{subsectioncolor}
\setlength{\arrayrulewidth}{1pt}
{\sffamily\bfseries\begin{tabular}{lp{6.75in}}\hline
\rowcolor{abstractbg}\multicolumn{2}{l}{\color{subsectioncolor}{\itshape
Highlights}{\Huge\strut}}\\
\rowcolor{abstractbg}$\bullet$ & We develop a 3D NSI framework for high-resolution volumetric ultrasound, outlining symmetry and element balance requirements, demonstrated with circular and spiral apertures at target acquisition volume rates.
\\
\rowcolor{abstractbg}$\bullet${\large\strut} & The proposed 3D NSI improves azimuthal and elevational resolutions by 36\% and contrast ratio by 24\%  compared to conventional delay and sum  without compromising computational efficiency. 
 \\
\rowcolor{abstractbg}$\bullet${\large\strut} & The new design of 3D NSI, featured with versatile apertures and computational efficiency, provides a practical solution for high-quality, high volume-rate ultrasound imaging in applications of depicting fast tissue dynamics.

 \\[2em]\hline
\end{tabular}}
\setlength{\arrayrulewidth}{0.4pt}
\arrayrulecolor{black}
\end{table*}

\bibliographystyle{ieeetr}

\section{Introduction}
\label{sec:introduction}
\IEEEPARstart{V}{olumetric} ultrasound imaging provides comprehensive anatomical and functional information for applications spanning cardiology, obstetrics, and interventional procedures. 
The pursuit of rich temporal information to capture dynamic physiological processes has been a major driver in developing ultrafast volumetric imaging techniques \cite{provost20143d}.
Like two-dimensional (2D) ultrasound, volumetric ultrasound faces a trade-off between spatial resolution and temporal resolution, but this challenge is further amplified in ultrasound systems with matrix arrays.

Achieving high azimuthal and elevational resolutions requires a large aperture with elements densely arranged in two dimensions at a proper pitch to mitigate spatial aliasing and prevent the appearance of grating lobes within the usable field of view. That the element pitch is typically bounded by half a wavelength often yields matrix arrays with thousands of elements, increasing both hardware complexity and data throughput demands.
This hardware requirement poses a practical challenge because typical commercial ultrasound systems have far fewer independent transmit/receive channels (e.g., 128-256) than the number of elements in modern matrix arrays (e.g., 1024). The mismatch between channel and element counts makes it impossible to address all elements of a matrix array simultaneously. Hence, low channel-count systems resort to multiplexing (MUX) to time-share channels among element groups in a large, densely-populated matrix array by sequentially addressing subsets of the array aperture to synthesize full-aperture coverage \cite{bernal2020high}. Despite its effectiveness in full-aperture synthesis, multiplexing significantly reduces acquisition volume rates, making real-time imaging of fast dynamics, such as cardiac motion, blood flow, and tracking of interventional tools, more difficult. 

To mitigate these constraints, several complementary strategies have been explored. 
Micro-beamforming (or, subarray beamforming) integrates analog beamforming inside the probe to fully exploit addressable elements in dense matrix arrays without a proportional increase in the system channel count\cite{larson19932, savord2003fully, matrone2014volumetric}.
In terms of array design, sparse arrays reduce the active element count while achieving image quality comparable to dense arrays by optimizing element arrangement and apodization across different pitches\cite{lookwood2002optimizing, austeng2002sparse, davidsen1994two, lockwood2002real}. This category includes designs ranging from various sparse matrix arrays \cite{jacquet2025simulation, roux2018experimental, 8832242, ramalli2015density} to specialized configurations, like Costas sparse 2-D arrays \cite{masoumi2023costas}. 
Another efficient design is the row-column arrays (RCA), which consist of two orthogonal one-dimensional (1D) arrays with electrical connections to either rows or columns, present a large aperture with a reduced number of elements that is compatible with low channel-count systems, thereby enabling high-volume-rate acquisitions \cite{rasmussen20153, seo2009256, savoia2007p2b}.

Beyond hardware, innovations in transmission sequences further push the limits of volumetric imaging. Techniques, such as employing diverging waves with a sparse virtual source design \cite{provost20143d, optvirtual, chen2017feasibility, wurowtrans, wodnicki2025handheld}, enable high-speed acquisitions by insonifying a large volume with a small number of transmissions. Moreover, advanced 3D ultrasound imaging has been applied to the aforementioned 2D array configurations to depict microvessels with RCA \cite{11124235}, 2D matrix array \cite{11082392}, 2D sparse array\cite{harput20193}, and 2D large pith array\cite{haidour2025multi}.
While these advancements in array design and transmission sequences address data acquisition challenges, they do not fully resolve the fundamental limitations of lateral or azimuthal resolution inherent to ultrasound imaging, prompting researchers to explore adaptive beamforming techniques originally developed for focused ultrasound systems. Techniques, such as Generalized Coherence Factor (GCF) \cite{li2003adaptive}, Minimum Variance (MV) \cite{synnevag2007adaptive}, Delay Multiply and Sum (DMAS) \cite{matrone2014delay}, and p-th Root Delay and Sum (p-DAS) \cite{polichetti2018nonlinear}, have been applied to 2D imaging. For 3D imaging, methods like Directional Coherence Factor (DCF) \cite{wucoherence}, sparse convolutional beamformer \cite{cohen2021sparse}, adaptive weight-based coherence beamformer \cite{yan2023fast}, and reinforcement learning-based beamforming using sparse spiral 2D arrays with non-linear beamformers (CF, DMAS-CF) \cite{10965879} have been proposed. All these beamformers can improve spatial resolution. Although effective, these methods often impose considerable computational overhead, complicating real-time volumetric operation at high volume rates.

Null Subtraction Imaging (NSI) is a nonlinear beamforming approach that balances between spatial resolution and computational complexity. It enhances lateral resolution of 2D ultrasound images using three different apodizations while keeping computational cost close to DAS (approximately 3×) \cite{agarwal2018improving, kou2022grating, 10918915, yociss2021null}. Motivated by its favorable balance between image quality and efficiency, we propose extending NSI to volumetric imaging, jointly addressing the hardware limitations inherent in a matrix array configured for multiplexed connection to an ultrasound system.


This work introduces a 3D NSI framework demonstrated on a 1024-element matrix array probe connected to a 256-channel ultrasound system.
Our contributions are threefold:

1. A 3D NSI framework: We develop a comprehensive theoretical framework for the NSI beamformer tailored for volumetric ultrasound. Our formulation identifies critical requirements for symmetry and balanced element counts in the receive aperture to produce high-resolution beampatterns. The proposed principles are demonstrated first through implementations using a fully-addressed circular aperture and a Fermat's spiral aperture, with applicability to rectangular apertures.

2. Hardware-aware apodization and high volume rates: We further propose a spiral 'no-reuse' design that enforces non-overlapping element sets across multiplexed banks. The no-reuse configuration eliminates multiplexing conflicts and is therefore a multiplexing-aware sparse aperture design, thereby increasing acquisition volume rates.

3. Mitigation of the resolution-speed trade-off: We present simulation and phantom results that evidence simultaneous improvements in resolution and contrast over DAS with low computational overhead (\textless 3$\times$ DAS), along with an analysis of volume-rate gains aligned with realistic channel count and MUX constraints.

The remainder of this paper is organized as follows. Section II describes the theory and design of 3D NSI based on a circular aperture and Fermat's spiral layouts with and without re-used elements between banks. They are followed by the implementation of transmission-reception (TX-RX) sequences, computer simulation, and experimental setup. Section III presents the simulation and experimental results and comparison across methods. Section IV discusses the results and their implications and suggests future research directions. Finally, Section V concludes the work and highlights its potential clinical applications.

\section{Methods}

\subsection{Design of 2D Aperture and Apodization}

The selection of active elements is based on the gridded layout of a commercial 2D matrix array probe (Vermon S.A., Tours, France). It has a center frequency of 3.5 MHz, a pitch of 300 $\mu$m, and a 70\% fractional bandwidth. It consists of 32 columns ($x$-axis; azimuth) and 35 rows ($y$-axis; elevation), where the 9th, 17th, and 25th are blank rows for electrical wiring. In total, there are 32 active rows, yielding 32 × 32 = 1024 addressable elements. 
The array is interfaced to a Verasonics Vantage system through a 4-to-1 multiplexer (UTA 1024-MUX, Verasonics Inc.), which partitions the aperture into four banks of 32$\times$8 elements. In our setup, the system provides 256 parallel channels per acquisition, necessitating sequential bank selection to synthesize the full 2D aperture.


In this study, we consider three aperture configurations (or, element masks) for volumetric imaging: a fully-addressed circular aperture, a sparse Fermat’s spiral aperture, and a spiral no-reuse aperture specially designed to resolve multiplexing conflicts. Each aperture is paired with a corresponding set of NSI apodization weights (see section II-B). 

\begin{figure}[!t]
\centerline{\includegraphics[width=\columnwidth]{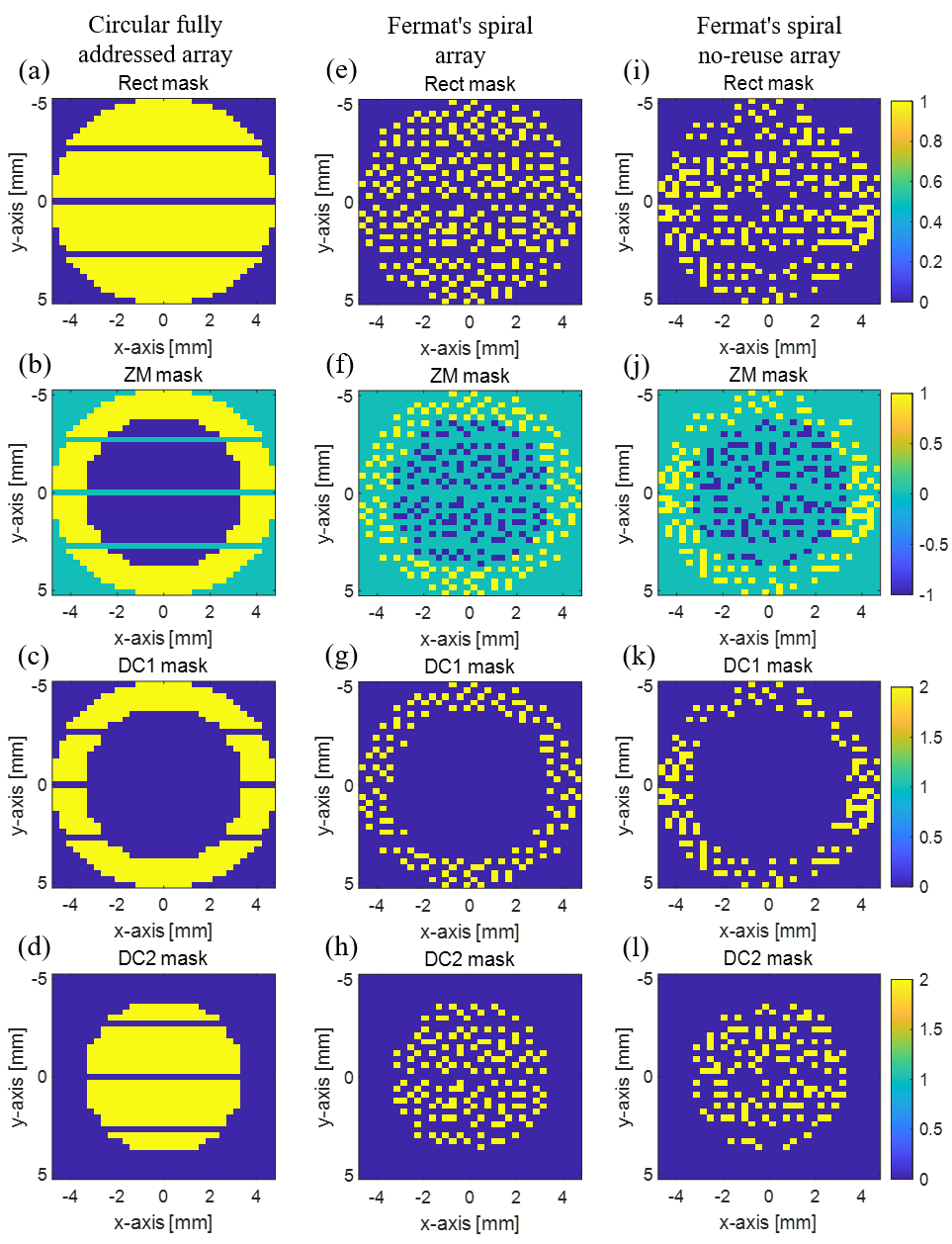}}
\caption{Apodization weights used in 3D DAS and 3D NSI imaging in the fully-addressed circular (a-d), Fermat's spiral (e-h), and spiral no-reuse (i-l) configurations. (a, e, i): Rectangle 
apodization for DAS; (b, f, j): ZM apodization for NSI; (c, g, k): DC1 apodization for NSI; (d, h, l): DC2 apodization for NSI; DAS: delay and sum; NSI: null subtraction imaging.}
\label{fig1}
\end{figure}

The circular aperture, chosen for its theoretical isotropic beampattern and aperture size comparable to the spiral designs, serves as a performance benchmark.
All elements within a specified circular radius centered in the array are activated, subject to hardware bank partitioning. The aperture diameter is defined to maximize usable aperture with a coaxial design as shown in Fig. \ref{fig1}(a). 


The sparse spiral array design is based on the Fermat's spiral layout according to the method described in \cite{ramalli2015density} to relax the channel count. In this study, 256 ideal positions were generated within the circular aperture. Each ideal position was then quantized to the closest physical element location on the Vermon array grid, yielding a sparse spiral selection mapped to the matrix array (Fig. \ref{fig1}(e)).
Because multiplexing time-shares electronic channels across the four banks for elements with identical $x$–$y$ indices, this spiral selection includes repeated use of channel indices across banks. This sparse spiral array still requests bank-by-bank selections and produces the same volume rate as the fully-addressed circular array.

To address the element re-use issue associated with the aforementioned spiral sparse array and to achieve a higher volume rate with current matrix probe setting, we propose to select elements that do not share the same channel. 
This selection enables higher volume rates by adapting the Fermat's spiral configuration and is referred to as the "spiral no-reuse" in this study (Fig. \ref{fig1}(i)). To determine which element to use, we propose a selection score $S_{final}(i,j)$ for each candidate physical transducer element $j$ relative to each ideal point $i$ using a Gaussian kernel as follows:
\begin{equation}
    \label{eq3}
    S_{final}(i,j) = exp(-\frac{d_{min}(i,j)}{2\sigma^2_d})
\end{equation}  
where $d_{min}(i,j)$ is the minimum distance between each physical candidate element location and an ideal location, $\sigma_d$ denotes the standard deviation of the Gaussian kernel used to score the element distance, $\sigma_d =0.7$ is used in this study.
For each of the local coordinate sets, we compare the $S_{final}$ scores of the candidate transducer elements across the four physical banks. The element with the highest score across the banks was uniquely selected, resulting in a conflict-free matrix configuration. This led to a total of 240 elements across all four banks that can be operated without multiplexing (Fig. \ref{fig1}(i)). This design enables a 16-fold increase in the maximum achievable volume rate compared to the bank-by-bank TX-RX used in both cases of fully-addressed circular aperture and spiral sparse aperture,where multiplexing is required due to channel sharing. Table \ref{tabhard} summarized the key parameters of the three configurations used in this study.
\begin{table}
\caption{PARAMETER COMPARISON ACROSS THREE ARRAY CONFIGURATIONS}
\renewcommand\arraystretch{1.2}
\label{table}
\setlength{\tabcolsep}{3pt}
    \begin{tabular}{m{30pt}<{\centering}|m{39pt}<{\centering}|m{39pt}<{\centering}|m{39pt}<{\centering}|m{39pt}<{\centering}|m{35pt}<{\centering}}
    \hline
       & \# of active elements & \# of TX/RX events per angle & total \# of events used per compounded volume & RF data size per volume & Maximum volume rate  \\ \hline
       Circ  & 812 & 16 & 144 & 71MB & 76 \\ \hline
        Spiral  & 256 & 16 & 144 & 22MB & 76 \\ \hline
        Spiral no-reuse & 240 & 1 & 9 & 21MB & 1222  \\ \hline
    \end{tabular}
\label{tabhard}
\end{table}

\subsection{3D Null Subtraction Imaging (NSI)}
Null subtraction imaging utilizes three apodization windows on the receive aperture, including a zero-mean (ZM) window and two windows with a DC offset on the ZM window, to reduce side lobes and grating lobes while narrowing the main lobe width as described in prior NSI works\cite{kou2022grating, agarwal2018improving, kou2023thesis}.

Here, we outline 3D NSI for single-angle diverging wave transmission: 

\textit{Step 1}. Apodization Window Generation: Three distinct apodization masks are created for a 2D array. The ZM apodization consists of weights with equal numbers of +1 and -1 on the outer and inner halves of the receive aperture, respectively, and is defined as
\begin{equation}
    \label{eq10}
    A_{ZM,i} = \left\{ \begin{matrix}
    {-1,~~0 \leq r_i \leq r_{in}~} \\
    {1,~~r_{in} < i \leq r_{out}}
\end{matrix} \right.,
\end{equation}
where $A_{ZM,i}$ is the zero-mean apodization weight for element $i$, $r_i$ is the distance between the element $i$ and the center of the array, $r_{out}$ is the outer radius of the circular aperture, and $r_{in}$ is the inner radius. To achieve coaxial symmetry between the inner and outer apertures, the total number of elements used in each should be approximately equal in order to ensure equal energy.

The second apodization adds a constant DC offset, $dc$, to $A_{ZM,i}$ and is defined as
\begin{equation}
    \label{eq11}
    A_{DC1,i} = A_{ZM,i} + dc,
\end{equation}
and the third apodization is an inner-outer flip of the second apodization and defined as 
\begin{equation}
    \label{eq11}
    A_{DC2,i} = -A_{ZM,i} + dc.
\end{equation}

\textit{Step 2}. Parallel Beamforming and Envelope Detection: Acquired radio-frequency (RF) data are independently beamformed three times with standard DAS, each time using a different apodization mask ($A_{ZM},A_{DC1}, A_{DC2}$). Receive beamforming is followed by envelope detection, producing corresponding envelope data  ($E_{ZM},E_{DC1}, E_{DC2}$).

\begin{figure}[!t]
\centerline{\includegraphics[width=\columnwidth]{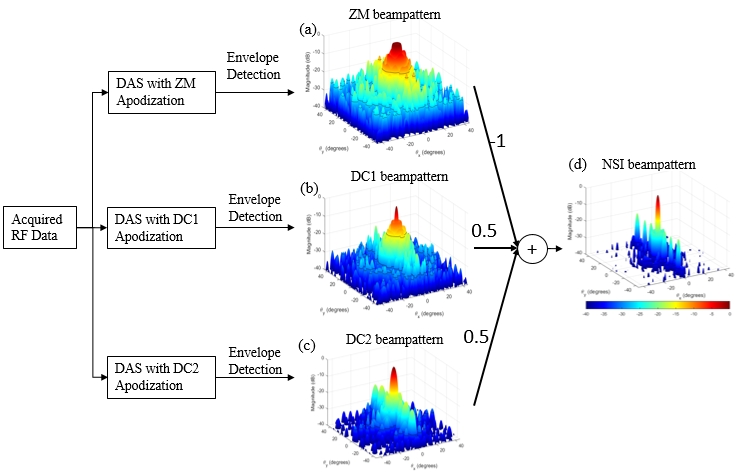}}
\caption{Block diagram of 3D NSI: Surface plots of the beampatterns of (a) ZM  (b) DC1, (c) DC2 apodizations and (d) NSI.}
\label{fig2}
\end{figure}

\textit{Step 3}. Nonlinear Image Combination: The final high-resolution NSI image ($E_{NSI}$) is synthesized by combining the three envelope data. The core principle is to subtract the ZM image ($E_{ZM}$), which has a null at the center of its beampattern, from the average of the two DC offset images ($E_{DC1}$ and $E_{DC2}$). The NSI image is subsequently normalized, log compressed, and displayed in gray scale. 
The general block diagram of 3D NSI is shown in Fig. \ref{fig2}.

\begin{figure}[!t]
\centerline{\includegraphics[width=\columnwidth]{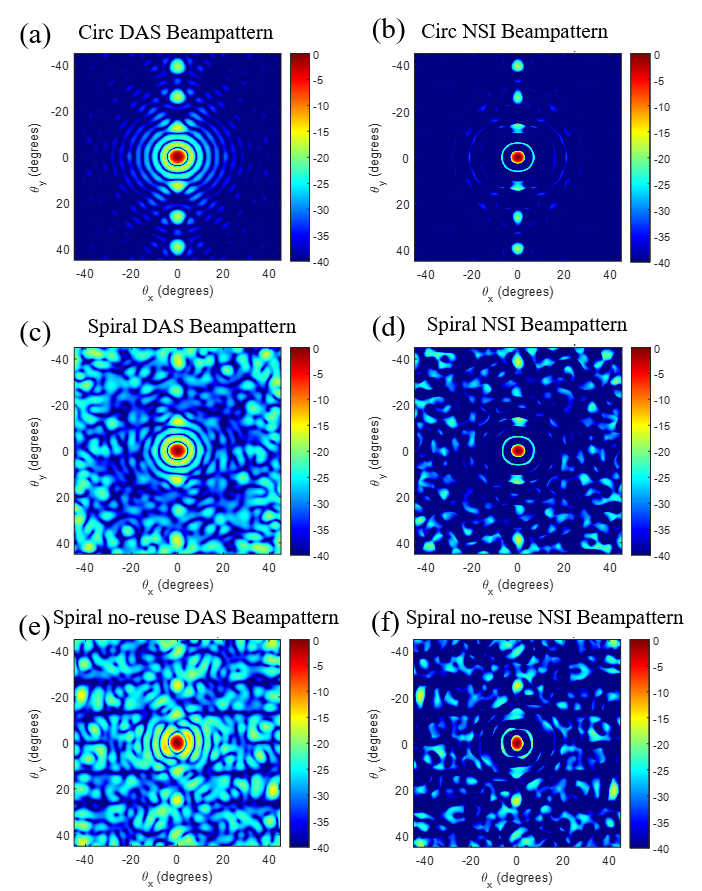}}
\caption{Comparison of beampatterns between 3D DAS (a, c, e) and 3D NSI (b, d, f) using circular (a, b), spiral (c, d), and spiral no-reuse (e, f) apodizations.}
\label{fig3}
\end{figure}

To balance the number of active elements between the inner and outer regions, $r_{in}$ and $r_{out}$ were calculated first based on the $32\times32$ elements square mask.
We set $r_{out}=16\times$ pitch and $r_{in}=11.5\times$pitch, where pitch is the element spacing. Then, the elements are mapped to the four $8\times32$-element banks. Fig. \ref{fig1} shows the design of NSI apodization windows accordingly. For the fully-addressed circular aperture, there were 408 inner and 404 outer elements (Fig. \ref{fig1}(a-d)); for the spiral configuration, the total numbers of inner and out elements were 132 and 124, respectively (Fig. \ref{fig1}(e-h)); the spiral no-reuse array had 118 inner elements and 122 outer elements (Fig. \ref{fig1}(i-l)). Thus, all three configurations had a roughly equal number of elements used in the inner and outer areas. The theoretical beampatterns of the three apodized apertures shown in Fig. \ref{fig3} demonstrated that NSI (Fig. \ref{fig3} (b, d, f)) produced a narrower main lobe in both azimuthal and elevational directions compared to DAS (Fig. \ref{fig3} (a, c, e)).

\subsection{Simulation and Experimental setup}
\subsubsection{Simulations}
Computer simulations were carried out using Field II \cite{jensen1997field} and Matlab 2022b (The MathWorks, Natick, MA, USA). The ultrasound image simulations considered the matrix array probe described in section II.A, operating at 3 MHz.

Two numeric phantoms were created. The first consisted of five point scatterers in water, placed on the axis of the probe at depths of 20, 30, 40, 50, and 60 mm, respectively, to evaluate the point spread functions (PSFs) at different depths. 

The second numeric phantom was a tissue-mimicking phantom containing a background tissue with randomly distributed scatterers whose scattering amplitudes followed a zero-mean Gaussian distribution. 
To simulate fully developed speckle, each resolution cell had approximately 20 scatterers.
The phantom was centered at a depth of 40 mm, with a volume of 40 mm $\times$ 40 mm $\times$ 30 mm, respectively, in the $x$, $y$, and $z$ directions. Furthermore, a 10-mm diameter anechoic cyst was centered at (0, 0, 40) mm with its scattering amplitude set to one-fifth that of the surrounding tissue to assess the image contrast performance.

\begin{figure}[!t]
\centerline{\includegraphics[width=200pt]{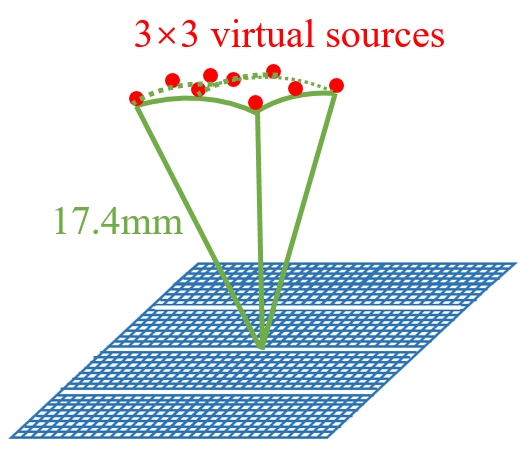}}
\caption{Illustration of the 9 virtual sources used to transmit diverging waves with matrix array. The virtual sources are located behind the array at a distance of 17.4 mm from the array center.}
\label{fig4}
\end{figure}

We used a diverging wave sequence consisting of nine transmission angles from their corresponding nine virtual point sources (Fig. \ref{fig4}). The virtual point sources were positioned at a total distance of 17.4 mm behind the array from the center of the array. 
The first transmission was directed straight ahead (0°, 0°) with the source positioned to face the center of the array. The remaining eight transmissions were steered to azimuth-elevation angle pairs of ($\pm$5°, 0°), (0°, $\pm$5°), and ($\pm$5°, $\pm$5°). These together formed a 3×3 angular grid as shown in Fig. \ref{fig4}. The ±5° increments provided a modest balance between the sector coverage and the volume rate for spatial compounding in ultrafast imaging. NSI with DC = 1 was used.

\subsubsection{Phantom Experiments}
Two separate phantom experiments were conducted for image quality evaluation using a CIRS ATS Model 539 phantom and a homemade phantom with four polyethylene wire targets in water. The first set of experiments used the commercial phantom, where point targets and anechoic cyst targets were scanned to assess image quality, including spatial resolution and contrast. The wire phantom was used to evaluate the side lobe level.

In both phantom experiments, the same diverging wave imaging sequence and hardware were used to acquire raw RF data as described in the simulation subsection and section II.A, respectively, with the matrix array probe driven at 15 volts.
For the fully-addressed circular and sparse spiral configurations, the process for a given steering angle is as follows: The matrix array probe used in this study consists of four banks. Each bank transmitted diverging waves four times. Following each transmission, a different bank received echoes. This resulted in four receive events per bank. Therefore, there were a total of 16 transmit-receive events—equivalent to the data acquired by a fully-addressed array. In contrast, the sparse no-reuse configuration required only a single transmission and a corresponding reception per steering angle.

\subsection{Performance Evaluation}

For each point/wire target, the 3-D beam profile passing through the maximum signal amplitude was obtained. Spatial resolution was quantified by measuring the full width at half maximum (FWHM) from the profile. 
Side-to-main lobe energy ratio (SMER)\cite{ramalli2015density} was used to evaluate side lobe levels, and it was calculated as the ratio of the summed amplitude of the sidelobes from -40 dB to -6 dB to the summed amplitude of the main lobe from -6 to 0 dB as follows:
\begin{equation}
    \label{eq12}
     SMER = 20\cdot log_{10}\frac{\int_{-x_{-40dB}}^{-x_{-6dB}}I(x)+\int_{x_{-6dB}}^{x_{-40dB}}I(x)}{\int_{-x_{-6dB}}^{x_{-6dB}}I(x)}
\end{equation}
 
Metrics adopted in this study for ultrasound image contrast evaluation include contrast ratio (CR) and contrast-to-noise ratio (CNR). The definitions of CR and CNR are as follows:
\begin{equation}
    \label{eq13}
    CR = ~\frac{\mu_{o} - \mu_{i}}{\mu_{o} + \mu_{i}},
\end{equation}

\begin{equation}
    \label{eq14}
    CNR = ~\frac{\mu_{o} - \mu_{i}}{\sqrt{\sigma_{i}^{2} + \sigma_{o}^{2}}},
\end{equation}
where $\mu_i$ and $\mu_o$ represent the mean amplitudes of the regions inside and outside the target, and $\sigma_i$ and $\sigma_o$ denote their corresponding standard deviations.

\section{Results}

\subsection{Simulations}

\begin{figure}[!t]
\centerline{\includegraphics[width=\columnwidth]{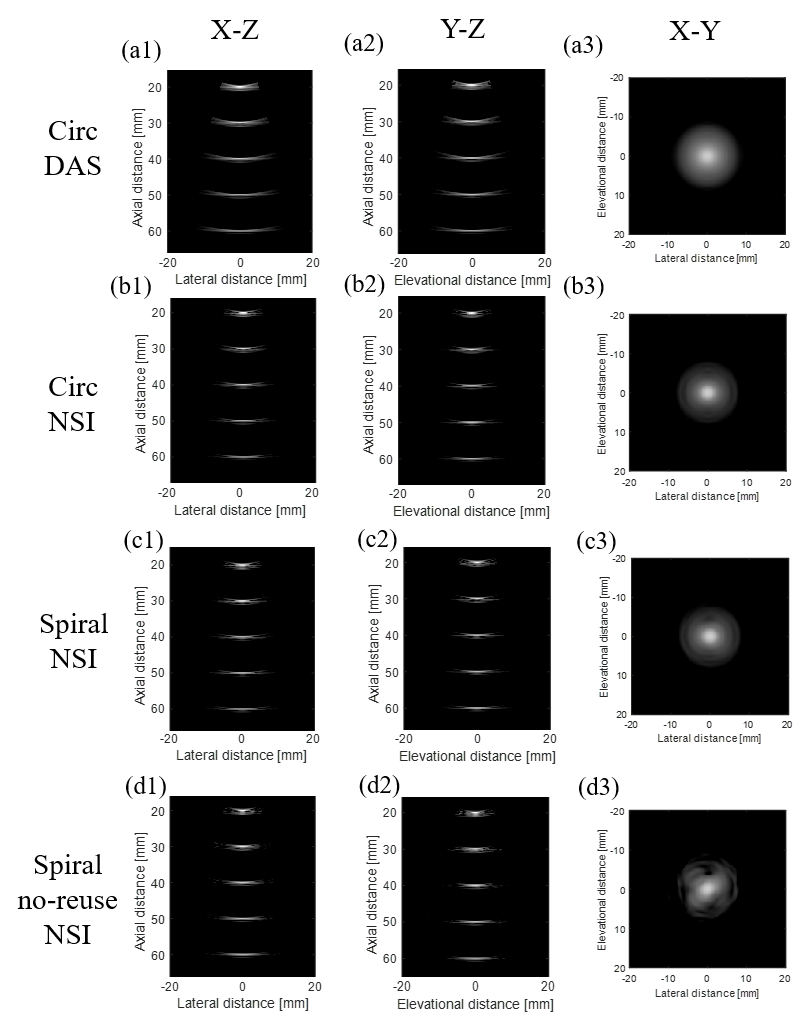}}
\caption{B-mode images of simulated point targets in the azimuth-depth plane (column 1), elevation-depth plane (column 2), and top (column 3, depth = 40 mm) views by (a1)–(a3) circular DAS, (b1)–(b3) circular NSI, (c1)–(c3) spiral NSI, and (d1)–(d3) spiral no-reuse NSI. Dynamic range is 50 dB for all images.}
\label{fig_simu_points}
\end{figure}

\begin{figure}[!t]
\centerline{\includegraphics[width=\columnwidth]{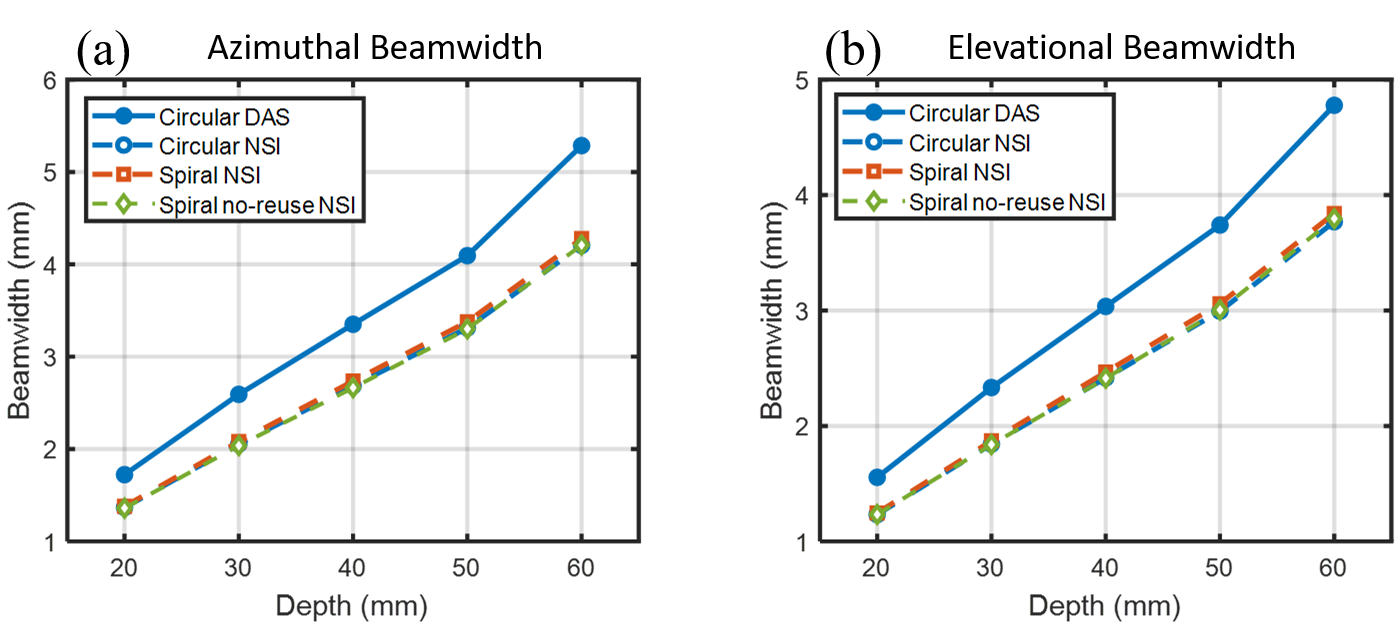}}
\caption{Comparison of beamwidths at five simulated point targets: FWHM measurements in (a) azimuth and (b) elevation for DAS and 3D NSI with circular, spiral, and spiral no-reuse apertures.}
\label{fig_simu_res}
\end{figure}

Figure \ref{fig_simu_points} shows B-mode images of five simulated point targets in three orthogonal view planes  in the three aperture configurations with 3D NSI as well as the circular aperture with DAS . The measured beamwidths (Fig. \ref{fig_simu_res}) consistently showed a narrower main lobe in 3D NSI compared to DAS across all three aperture configurations. Table \ref{tab:simu_res} reported that 3D NSI on average reduced the azimuthal beamwidth by $20\% \pm 0.7\%$ and elevational beamwidth by $ 20.5\% \pm 0.5\% $ and in total reduced around 36\% beam area across different depths.

Table \ref{tab:simu_res} also summarized the SMERs of the simulated point target at 40 mm depth in all imaging configurations for evaluating the capability of 3D NSI in side lobe suppression. Overall, 3D NSI achieved a side lobe level that was 2 to 3 dB lower than DAS regardless of the array configurations.
 



\begin{table}\centering
\caption{Comparison of Resolution and SMER: Performance of DAS and 3D NSI across Circular and Spiral Apertures in the Simulation Study at Depth = 40 mm}
\renewcommand\arraystretch{1.2}
\label{tab:simu_res}
\setlength{\tabcolsep}{4pt}
\begin{tabular}{m{45pt}<{\centering}|m{40pt}<{\centering}|m{40pt}<{\centering}|m{40pt}<{\centering}|m{45pt}<{\centering}}
\hline
Metric & Circ DAS & Circ NSI & Spiral NSI & Spiral no-reuse NSI \\ \hline
Lat FWHM (mm) & 3.35 & 2.68 & 2.74 & 2.66 \\ \hline
Ele FWHM (mm) & 3.04 & 2.41 & 2.47 & 2.41 \\ \hline
Lat SMER (dB) & -16.14 & -18.64 & -18.35 & -18.71 \\ \hline
Ele SMER (dB) & -16.42 & -19.26 & -19.03 & -19.24 \\ \hline

\end{tabular}
\end{table}

\begin{figure}[!t]
\centerline{\includegraphics[width=\columnwidth]{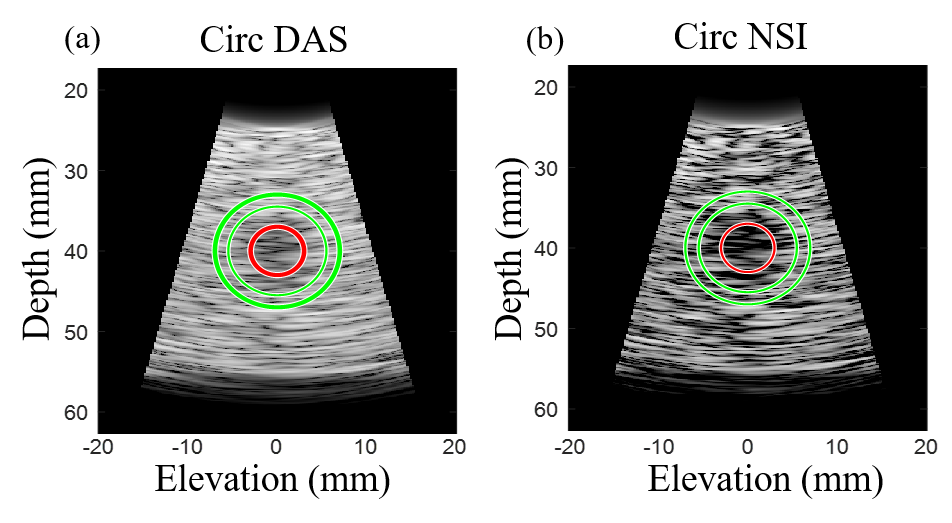}}
\caption{Contrast analysis in the cyst simulation study: B-mode images obtained with (a) circular DAS (b) circular NSI. The area between the two green circles was chosen as the background signal region, while the anechoic region is enclosed by the red circle. Images are displayed in a dynamic range of 40 dB.}
\label{fig_simu_contrast}
\end{figure}

The CR and GCNR are summarized in Table \ref{tab2}. They were calculated from the anechoic cyst and the background region in both X-Z and Y-Z planes as shown in Fig. \ref{fig_simu_contrast}.
The results demonstrated that 3D NSI consistently achieved a higher CR than DAS across all apodization schemes. For example, with the circular aperture, NSI increased CR in the X-Z plane from 0.47 (DAS) to 0.61, representing a 29\% improvement.
\begin{table}
\caption{Comparison of Contrast Between DAS and 3D NSI across Circular and Spiral Apertures with DAS/3D NSI in the Simulation Study}
\renewcommand\arraystretch{1.2}
\label{table}
\setlength{\tabcolsep}{3pt}
    \begin{tabular}{m{25pt}<{\centering}|m{30pt}<{\centering}|m{30pt}<{\centering}|m{30pt}<{\centering}|m{30pt}<{\centering}|m{30pt}<{\centering}|m{30pt}<{\centering}}
    \hline
      
        & \multicolumn{2}{|c|}{Circ} & \multicolumn{2}{|c|}{Spiral} &  \multicolumn{2}{|c}{Spiral no-reuse} \\ 
        \hline
        X-Z plane& DAS & NSI & DAS & Spiral & DAS & NSI  \\ \hline

        CR  & $0.47$ & $0.61$ & $0.37$ & $0.47$ & $0.41$ &$0.49$\\ \hline
        CNR & $1.10$ & $0.83$ & $1.00$ & $0.74$&$1.09$& $0.75$  \\ \hline
        \hline
        Y-Z plane& DAS & NSI &  DAS & NSI & DAS & NSI  \\ \hline
        
        CR  & $0.43$ & $0.57$ & $0.42$ & $0.49$ & $0.40$ &$0.47$\\ \hline
        CNR & $1.03$ & $0.80$ & $0.95$ & $0.67$&$0.95$&$0.63$  \\ \hline
    \end{tabular}
\label{tab2}
\end{table}

\subsection{Phantom Experiments}

\begin{figure}[!t]
\centerline{\includegraphics[width=\columnwidth]{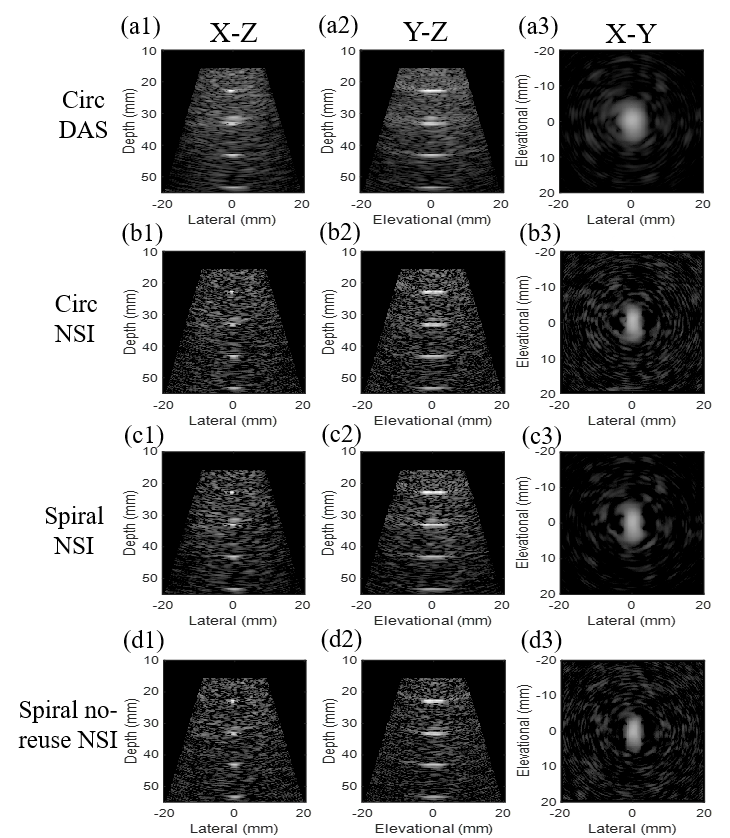}}
\caption{B-mode images of four wire targets in a CIRS ATS Model 539 multipurpose phantom in the cross-sectional (column 1), longitudinal (column 2), and top (column 3, depth = 44 mm) views by (a1)–(a3) circular with DAS, (b1)–(b3) circular with NSI, (c1)–(c3) spiral with NSI, and (d1)–(d3) spiral no-reuse with NSI configurations. The length direction of the wires were aligned with the elevational direction of the matrix array. Dynamic range is 50 dB for all images.}
\label{fig_phantom_lat_lines}
\end{figure}

Figure \ref{fig_phantom_lat_lines} compares the B-mode images of four wire targets in the CIRS phantom with their length direction perpendicular to the lateral–depth plane using the circular, spiral, and spiral no-reuse apertures with DAS and 3D NSI. 
The resolution and SMERs of the phantom line targets in all imaging configurations, summarized in Table \ref{tab:resolution}, showed a side lobe level 1 to 3 dB lower than DAS in the phantom experiments. Quantitative measures were derived based on 10 independent acquisitions of the same wire target.

\begin{figure}[!t]
\centerline{\includegraphics[width=\columnwidth]{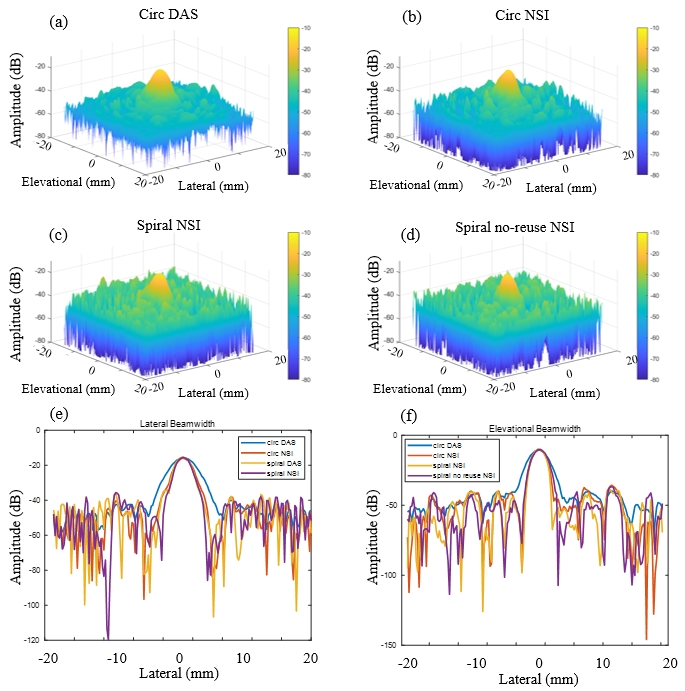}}
\caption{Surface plots of the beam patterns for a line target at depth of 55 mm in the same CIRS phantom imaged by (a) Circ DAS, (b) Circ NSI, (c) Spiral NSI, and (d) Spiral no-reuse NSI. One-dimensional beam profiles across the (e) azimuth and (f) elevation in the four aperture configurations.}
\label{fig_plc}
\end{figure}

\begin{figure}[!t]
\centerline{\includegraphics[width=\columnwidth]{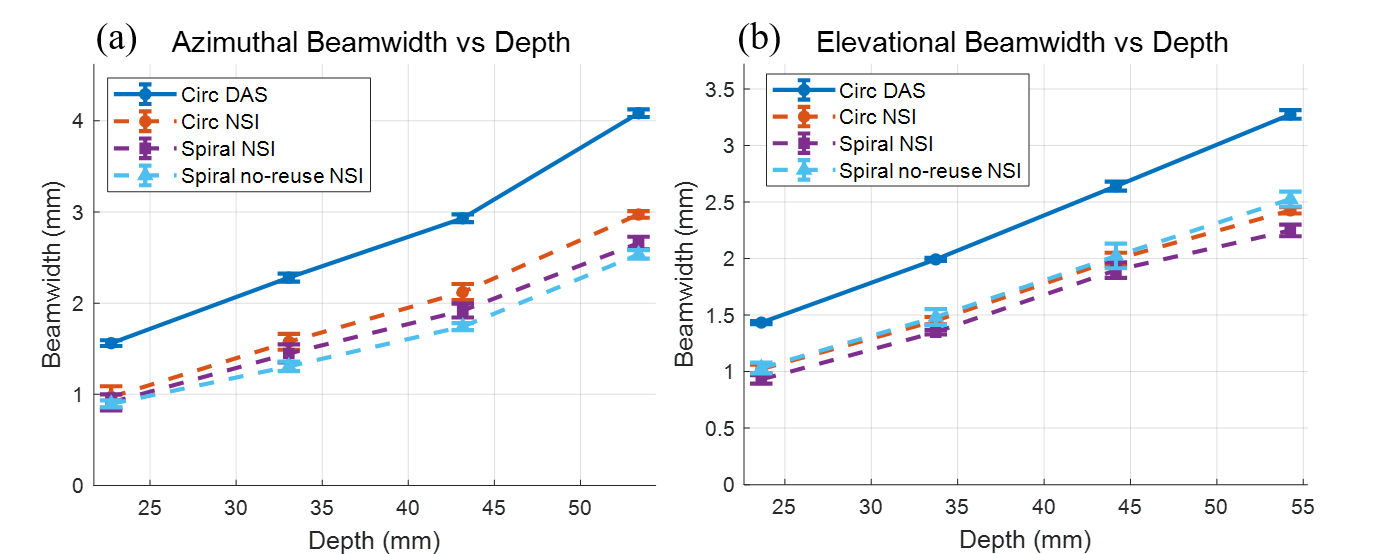}}
\caption{Comparison of beamwidths measured using FWHM of the line targets in the same CIRS phantom in the (a) lateral and (b) elevational directions for DAS and 3D NSI with circular, spiral, and spiral no-reuse apertures.}
\label{fig_phantom_latres}
\end{figure}

Fig. \ref{fig_plc} shows beam profiles across a line target at depth of 55 mm in the same CIRS phantom in the four aperture apodizations. As shown in Fig. \ref{fig_phantom_latres}, 3D NSI reduced the FWHM in both azimuthal and elevational directions across all aperture designs. At a depth of 44 mm, the resolution with the circular aperture improved from 2.69 mm (DAS) to 2.02 mm (3D NSI) in elevation and from 2.90 mm (DAS) to 2.05 mm (3D NSI) in azimuth, representing a 47\% reduction in the resolution area. Similar improvements were observed in Fig. \ref{fig_phantom_latres} for the spiral and spiral no-reuse configurations. These experimental results substantiated the resolution improvement observed in the simulations.

\begin{table}\centering
\caption{Comparison of Resolution and SMER metrics at Depth of 54 mm in the CIRS Phantom Experiment Across the Circular and Spiral Apertures with DAS/3D NSI}
\renewcommand\arraystretch{1.2}
\label{tab:resolution}
\setlength{\tabcolsep}{4pt}
\begin{tabular}{m{39pt}<{\centering}|m{43pt}<{\centering}|m{43pt}<{\centering}|m{43pt}<{\centering}|m{43pt}<{\centering}}
\hline
Metric & Circ DAS & Circ NSI & Spiral NSI & Spiral no-reuse NSI \\ \hline

Lat FWHM (mm) & $4.08 \pm0.04$ & $2.97 \pm0.04$ & $2.66 \pm0.07$ & $2.53\pm0.05$ \\ \hline
Ele FWHM (mm) & $3.27 \pm0.04$ & $2.43\pm0.03$ & $2.25\pm0.05$ & $2.53\pm0.06$ \\ \hline
Lat SMER (dB) & $-8.50 \pm 0.38$ & $-9.63 \pm0.37$ & $-10.30 \pm0.35$ & $-9.45\pm0.56$ \\ \hline
Ele SMER (dB) & $-8.78 \pm 0.14$ & $-10.15 \pm0.27$ & $-10.83 \pm0.41$ & $-11.47\pm0.26$ \\ \hline
\end{tabular}
\end{table}


\begin{figure}[!t]
\centerline{\includegraphics[width=\columnwidth]{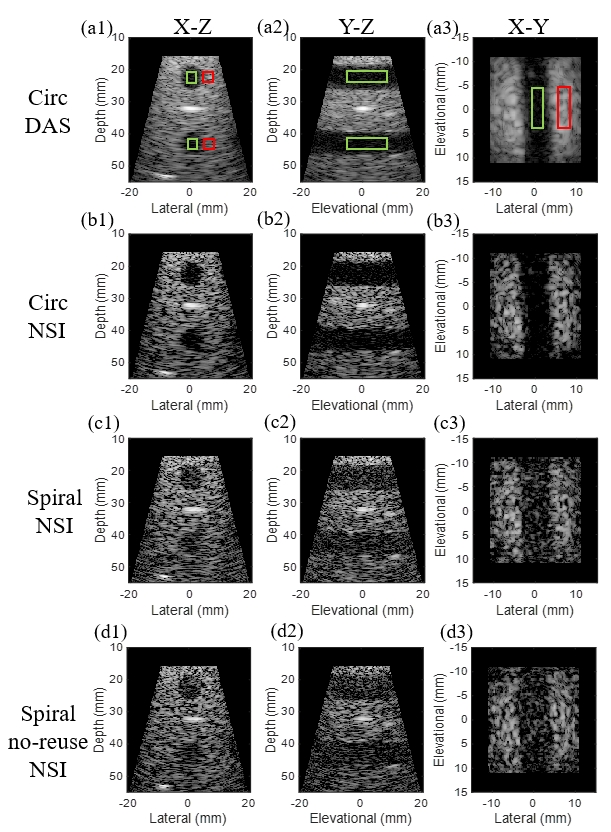}}
\caption{B-mode images of two 8-mm diameter cyst targets in the same CIRS phantom in the cross-sectional (column 1), longitudinal (column 2), and
top (column 3) views by (a1)–(a3) circular DAS, (b1)–(b3) circular NSI,(c1)–(c3) spiral NSI, and (d1)–(d3) spiral no-reuse NSI. Dynamic range is 50 dB for all images. The green and red boxes indicate the regions of interest for CR and CNR analysis.}
\label{fig_phantom_cyst}
\end{figure}

The green and red boxes in Fig. \ref{fig_phantom_cyst}(a1-a3) indicate the regions of interest that were used for the evaluation of contrast performance of the four imaging methods. Two cyst targets respectively located at depths of 20 mm and 45 mm in the CIRS phantom were assessed. Qualitatively, the fully-addressed circular aperture with either DAS or NSI exhibited higher contrast than the spiral cases; the spiral aperture with NSI had a better contrast than spiral no-re-use with NSI.  Quantitative contrast measurement, including CR and CNR, was obtained based on five independent acquisitions of the same cyst target and is summarized in Table \ref{tab3}.
For the cyst at 25 mm depth, 3D NSI achieved a CR of 0.92$\pm$0.01, surpassing 0.88$\pm$0.01 obtained with DAS in the case of the circular aperture; the spiral and spiral no-reuse NSI configurations improved the CR from 0.66$\pm$0.02 and 0.61$\pm$0.02 (DAS) to 0.72$\pm$0.02 and 0.69$\pm$0.01, respectively. These improvements in CR aligned with the qualitative observations of enhanced cyst boundary delineation and reduced side lobes shown in Fig. \ref{fig_phantom_cyst}. However, CNR was lower in the cases with NSI than with DAS.



\begin{table}
\caption{Contrast Analysis of Hypoechoic Regions in the CIRS ATS Model 539 Phantom Imaged with Different Methods}
\renewcommand\arraystretch{1.2}
\label{table}
\setlength{\tabcolsep}{3pt}
    \begin{tabular}{p{25pt}<{\centering}|p{30pt}<{\centering}|p{30pt}<{\centering}|p{30pt}<{\centering}|p{30pt}<{\centering}|p{30pt}<{\centering}|p{30pt}<{\centering}}
    \hline
           & \multicolumn{2}{|c|}{Circ} & \multicolumn{2}{|c|}{Spiral} &  \multicolumn{2}{|c}{Spiral no-reuse} \\ 
        \hline
       25mm cyst& DAS & NSI & DAS & NSI & DAS &  NSI  \\ \hline
        CR  & $0.88\pm0.01$ & $0.92\pm0.01$ & $0.66\pm0.02$ & $0.72\pm0.02$ & $0.61\pm0.02$ &$0.69\pm0.01$\\ \hline
        CNR & $1.73\pm0.04$ & $1.19\pm0.03$ & $1.44\pm0.03$ & $0.94\pm0.01$&$1.40\pm0.08$& $0.90\pm0.04$  \\ \hline
        \hline
       45mm cyst& Circ DAS & Circ NSI & Spiral DAS & Spiral NSI & Spiral no-reuse DAS & Spiral no-reuse NSI  \\ \hline
        CR  & $0.73\pm0.01$ & $0.78\pm0.03$ & $0.45\pm0.04$ & $0.57\pm0.06$ & $0.42\pm0.06$ &$0.54\pm0.07$\\ \hline
        CNR & $1.54\pm0.08$ & $1.11\pm0.10$ & $1.10\pm0.09$ & $0.80\pm0.08$&$1.04\pm0.10$&$0.78\pm0.09$  \\ \hline
    \end{tabular}
\label{tab3}
\end{table}

\section{Discussion}

This study introduces a 3D Null Subtraction Imaging (3D NSI) method that improves azimuthal and elevational resolutions while keeping computational cost low ($<$ 3× that of delay‑and‑sum). The method operates on beamformed data and is effective across diverse aperture geometries, including circular and spiral layouts. When integrated into a designed spiral no‑reuse aperture, the framework additionally increases the achievable volume rate by 16× without altering transmit sequences or sacrificing the resolution gains compared to the fully-addressed circular and spiral apertures.

\begin{figure}[!t]
\centerline{\includegraphics[width=\columnwidth]{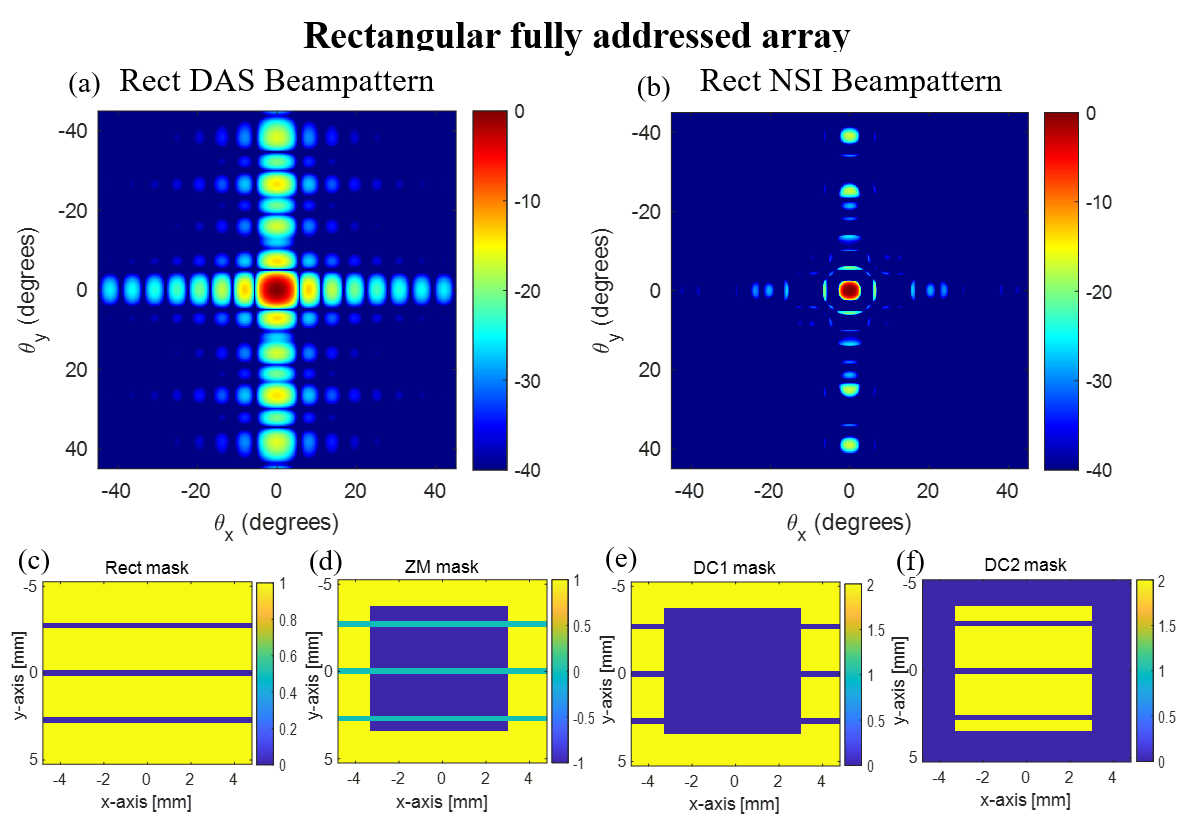}}
\caption{Comparison of beampatterns (top row) and apodizations (bottom row) used in (a, c) 3D DAS and (b, d-f) 3D NSI in the fully-addressed matrix array.}
\label{fig_rect}
\end{figure}

The results from both simulations and phantom experiments showed that 3D NSI consistently improved both azimuthal and elevational resolutions (an approximate 36\% reduction in resolution area), regardless of the aperture geometry being circular or spiral. This indicated that the proposed 3D NSI is not tied to a specific geometry. In other words, it is compatible with other aperture geometries, e.g., a rectangle, as shown in Figure \ref{fig_rect}, as long as a balanced and symmetric count of inner and outer elements is maintained (e.g., use of 22$\times$22 elements within the inner rectangular area) to ensure energy equilibrium.

A fundamental characteristic of 3D NSI is its function as a post‑processing step applied to received RF signals. This enables seamless integration into existing transmission and reception schemes regardless of wave type, transducer configuration, or pulse sequence. Like many nonlinear beamformers that modify speckle texture, NSI decreases CNR. Nevertheless, within the tested conditions, NSI markedly enhanced CR and visual depiction of tissue structures, facilitating image analysis tasks, such as lesion detection and delineation.

In this study, the exemplary 3D NSI framework assigned a value of +1 to the outer region and –1 to the inner region for the ZM window. It is important to note that the assignment of +1 and -1 is arbitrary and that this mapping can be reversed without significantly affecting the resultant NSI beam pattern.

From the perspective of computation, 3D NSI only involves linear operations on the beamformed data, resulting in a total processing overhead below 3× that of standard DAS. In contrast, coherence‑based weighting, minimum‑variance, or delay‑multiply‑and‑sum approaches typically introduce substantial additional computations during beamforming and are therefore challenging to scale up for real‑time 3D. Overall, the computational efficiency of 3D NSI improves the practicality of large volumes at ultrafast rates.


The study has several limitations. The spiral no-reuse aperture, although effective and adapted from the spiral configuration, may not represent the optimal active aperture design. Instead, it serves as a feasible option for using the matrix array without conflicts between elements in different banks. Element pitch, fixed layout, and element positions impose non-ideal spatial sampling. Using the average element positions in the matrix array as a physical constraint of the matrix array is not ideal.

A volume rate exceeding 1000 volumes per second at high resolution could facilitate 4D assessment of cardiac kinematics and hemodynamics, as well as the capture of transient cardiovascular (e.g., pulse wave) and neural (e.g., via cerebral blood flow) events. However, the practical maximum volume rate is constrained not only by algorithmic cost but also by front‑end bandwidth, memory I/O, GPU/CPU throughput, and thermal limits. These system‑level factors must be addressed to ensure sustained high volume-rate operation. 

Beyond the tested matrix array and aperture geometries, the presented 3D NSI approach is applicable to other 2D arrays with approximately symmetric azimuthal and elevational transducer responses. The observed suppression of side lobes and grating lobes suggests that larger‑pitch arrays may be used to achieve a larger field of view. 

Future work will focus on \textit{in vivo} validation to confirm efficacy in real biological tissues. 
Subsequent research trajectories could focus on a broader range of applications that rely on high resolution and frame rate, such as 3D power Doppler and ultrasound localization microscopy. Further technical developments may consider integrating coded excitation to mitigate the inherent SNR loss of NSI for deep tissue imaging as well as employing deep learning to optimize sparse aperture layouts tailored to specific clinical tasks.

\section{Conclusion}

This study presented and evaluated a 3D NSI framework for volumetric ultrasound imaging using a Vermon 1024‑element matrix array with a multiplexed connection to a 256-channel ultrasound system. Three aperture/apodization configurations were considered: fully-addressed circular , sparse spiral, and sparse spiral with no element reuse across banks. The results from computer simulations and quality assurance phantom experiments show that the proposed method consistently enhanced azimuthal and elevational resolutions, reduced side lobe energy, and improved contrast ratio in all three configurations. 
By resolving multiplexing conflicts, the spiral no-reuse configuration achieved a 16-fold increase in the acquisition volume rate. While the 3D NSI beamformer entails a computational cost of less than 3× that of DAS, this is a minor factor compared to the dominant savings in acquisition time, resulting in a significantly faster overall imaging pipeline.
By offering low computational complexity and improving image quality, 3D NSI with spiral no-reuse apodization is an effective and practical solution to overcome the current limitations of multiplexed matrix array imaging, potentially enabling real-time, high-resolution four-dimensional ultrasound applications.

\section*{Acknowledgment}

The authors would like to thank Dr. Xianchuan Wu for her help with the \textit{in vitro} experiments.
\bibliography{ref}

\end{document}